\documentclass{aastex63}

\usepackage{xspace}

\graphicspath{{./}{figures/}}

\submitjournal{PASP}

\newcommand{\Hubble}{{\it Hubble Space Telescope}\xspace}
\newcommand{\HST}{{\it HST}\xspace}
\newcommand{\RomanST}{{\it Roman}\xspace}
\newcommand{\RomanSpelled}{{\it Nancy Grace Roman Space Telescope}\xspace}
\newcommand{\xtimes}{\times\xspace}

\shorttitle{Forward Modeling for \RomanST SNe}
\shortauthors{Rubin and Cikota et al.}

\begin{document}

\newcommand{\uhawaii}{\affiliation{Department of Physics and Astronomy, University of Hawai`i at M{\=a}noa, Honolulu, Hawai`i 96822}}
\newcommand{\stsci}{\affiliation{Space Telescope Science Institute, 3700 San Martin Drive Baltimore, MD 21218, USA}}
\newcommand{\lbnl}{\affiliation{E.O. Lawrence Berkeley National Laboratory, 1 Cyclotron Rd., Berkeley, CA, 94720, USA}}

\author[0000-0001-5402-4647]{David Rubin}
\uhawaii
\affiliation{E.O. Lawrence Berkeley National Laboratory, 1 Cyclotron Rd., Berkeley, CA, 94720, USA}

\author[0000-0001-7101-9831]{Aleksandar Cikota}
\affiliation{E.O. Lawrence Berkeley National Laboratory, 1 Cyclotron Rd., Berkeley, CA, 94720, USA}
\affiliation{European Southern Observatory, Alonso de Cordova 3107, Vitacura, Casilla 19001, Santiago de Chile, Chile}

\author{Greg Aldering}
\lbnl

\author{Andy Fruchter}
\stsci

\author{Saul Perlmutter}
\lbnl
\affiliation{Department of Physics, University of California Berkeley, Berkeley, CA 94720, USA}

\author{Masao Sako}
\affiliation{Department of Physics and Astronomy, University of Pennsylvania, Philadelphia, PA 19104, USA}

\title{
Going Forward with the \RomanSpelled Transient Survey:\\ Validation of Precision Forward-Modeling Photometry for Undersampled Imaging}

\correspondingauthor{David Rubin}
\email{drubin@hawaii.edu}

\begin{abstract}

The \RomanSpelled (\RomanST) is an observatory for both wide-field observations and coronagraphy that is scheduled for launch in the mid 2020's. Part of the planned survey is a deep, cadenced field or fields that enable cosmological measurements with type Ia supernovae (SNe~Ia). With a pixel scale of $0\farcs11$, the Wide Field Instrument will be undersampled, presenting a difficulty for precisely subtracting the galaxy light underneath the SNe. We use simulated data to validate the ability of a forward-model code (such codes are frequently also called ``scene-modeling'' codes) to perform precision supernova photometry for the \RomanSpelled SN survey. Our simulation includes over 760,000 image cutouts around SNe~Ia or host galaxies ($\sim 10\%$ of a full-scale survey). To have a realistic 2D distribution of underlying galaxy light, we use the VELA simulated high-resolution images of galaxies. We run each set of cutouts through our forward-modeling code which automatically measures time-dependent SN fluxes. Given our assumed inputs of a perfect model of the instrument PSFs and calibration, we find biases at the millimagnitude level from this method in four red filters ($Y106$, $J129$, $H158$, and $F184$), easily meeting the 0.5\% \RomanST inter-filter calibration requirement for a cutting-edge measurement of cosmological parameters using SNe Ia. Simulated data in the bluer $Z087$ filter shows larger $\sim$~2--3 millimagnitude biases, also meeting this requirement, but with more room for improvement. Our forward-model code has been released on Zenodo.

\end{abstract}


\section{Introduction} \label{sec:intro}

The \RomanSpelled (\RomanST) is an observatory for both wide-field observations and coronagraphy that is scheduled for launch in the mid 2020's. The Wide Field Instrument (WFI) covers 0.281 square degrees and performs both imaging and low-resolution slitless spectroscopy. One of the primary science objectives of the \RomanST mission is to investigate the expansion history of the Universe using thousands of Type Ia Supernovae (SNe~Ia). Although the \RomanST supernova survey strategy is not yet finalized, the survey is planed to have two components: a $\sim 5$-day cadence multi-band imaging survey to discover transients and measure their light curves, and a spectroscopic component to classify a subset of the transients and measure redshifts. The cadenced observations would take place over a period of about two years (146 visits), with each visit rotating $\sim 5^{\circ}$ from the previous visit (two full rotations over two years) to keep the solar panels pointed at the Sun. 

\begin{figure}
    \centering
    \includegraphics[width = \textwidth]{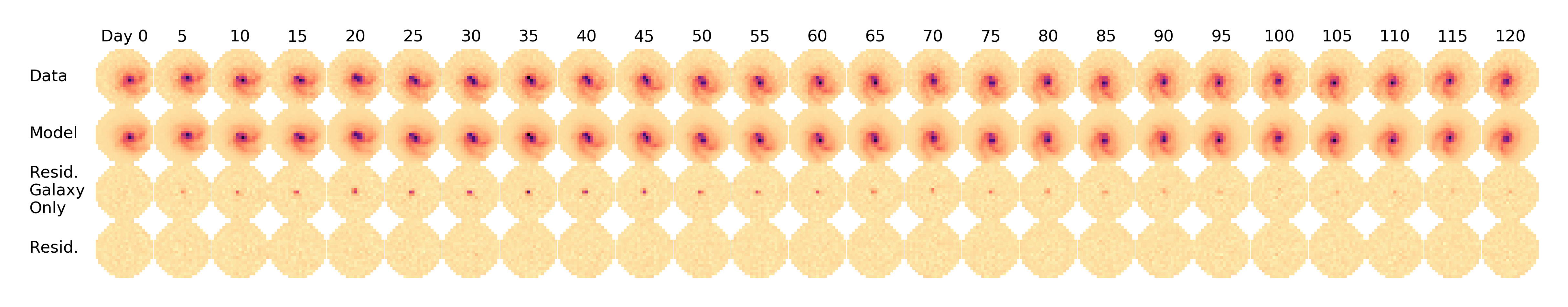}
    \caption{Forward modeling an example simulated SN in $Y106$. Each epoch ({\bf each column}) consists of only one undithered exposure per filter; subpixel sampling is provided by a $\sim 5^{\circ}$ rotation between subsequent epochs. The {\bf top row} of panels shows a set of cutouts around the SN; only epochs with SN light are shown. The {\bf next row} of panels shows the model inferred from this data and 49 reference epochs without the SN. The galaxy model is an analytic function, parameterized with a 2D set of spline nodes. For each image, the galaxy model is sampled at high resolution (11$\xtimes$ oversampling, across each pixel, then convolved with the PSF and the pixel and sampled at the native pixel scale. This accuracy is sufficient for $\sim 10^{-4}$ accuracy in representing the spline. The image-dependent SN light (also convolved by the PSF and the pixel and sampled at the native scale) and background are added to this galaxy model. The {\bf next row} of panels show the residuals where the SN models are not subtracted. Finally, the {\bf bottom row} of panels show the residuals when the modeled SN and modeled galaxy are subtracted. }
    \label{fig:forwardexample}
\end{figure}

\begin{figure}
    \centering
    \includegraphics[width=0.6\textwidth]{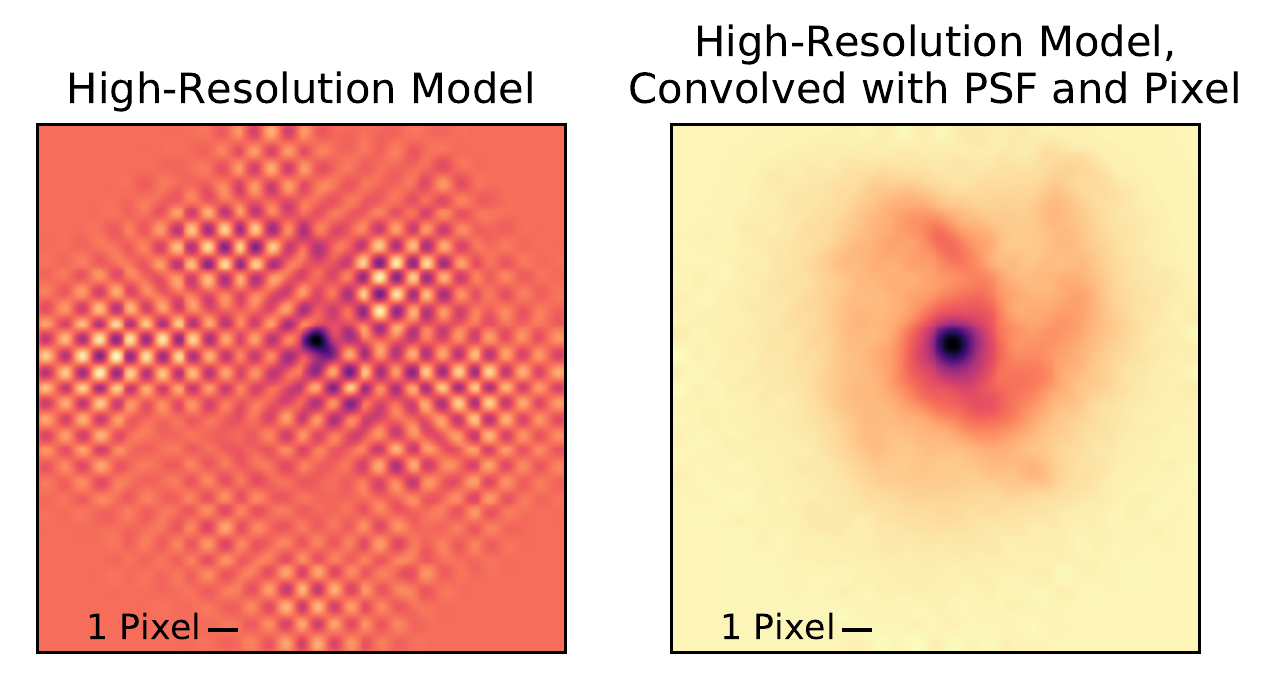}
    \caption{Recovered galaxy model from Figure~\ref{fig:forwardexample}. The {\bf left panel} shows the recovered high-resolution galaxy model $G$ (from Equation~\ref{eq:galaxymodel}), sampled at 11$\xtimes$ the native resolution (i.e., $0\farcs01$). The {\bf right panel} shows $G$ convolved with the PSF and the pixel (the PSF has the same orientation), also at 11$\xtimes$ the native resolution. Most of the unphysical high-frequency power visible in the left panel is suppressed. Note that the convolution with the pixel ensures that the total galaxy flux is not dependent on the  alignment with the pixels in a given epoch.
    \label{fig:forwardexamplehighres}}
\end{figure}

To balance field of view, read noise, and PSF sampling, the pixel scale of the WFI was set at $0\farcs11$, leaving the imaging PSF undersampled (Table~\ref{tab:filter}). This undersampling is not mitigated by the SN survey strategy, since much of the survey will likely only have one (undithered) exposure per filter per epoch in order to minimize overheads and read noise. Undersampled, undithered imaging poses a challenge for photometry methods based on image resampling (e.g., \citealt{alardlupton}). Much existing undersampled photometry is thus done with codes that use forward modeling (e.g., \citealt{suzuki12, hayden21}). Forward modeling (called ``scene modeling'' by \citealt{holtzman08}) bypasses image resampling to model each image as observed.\footnote{The ground-based SuperNova Legacy Survey developed a similar sort of code that used resampled and aligned images \citep{astier06, guy10}. The goal in that work was to avoid image subtraction and take the time-variable PSF into account.} Figure~\ref{fig:forwardexample} shows an example and Figure~\ref{fig:forwardexamplehighres} shows the recovered high-resolution galaxy model.

\begin{deluxetable}{cccccc}
\caption{Filter-dependent simulation quantities \label{tab:filter}}
\tablehead{
\colhead{Filter} & \colhead{$Z087$} & \colhead{$Y106$} & \colhead{$J129$} & \colhead{$H158$} & \colhead{$F184$}}
\startdata
Background (e$^-$/pix/s) & 0.349 & 0.384 & 0.376 & 0.365 & 0.381 \\
Exposure Time (s) & 300 & 300 & 300 & 300 & 600 \\
Read Noise (e$^-$) & 8.3 & 8.3 & 8.3 & 8.3 & 6.9 \\
Fitted PSF FWHM (") & $0\farcs 127$ & $0\farcs 130$ & $0\farcs 136$ & $0\farcs 150$ & $0\farcs 166$ \\
Gaussian Fitted PSF FWHM (Pixels) & 1.16 & 1.18 & 1.24 & 1.36 & 1.51 \\
\enddata
\end{deluxetable}

The gold standard for validating forward-model results is to inject simulated SNe into real survey data (\citealt{holtzman08, suzuki12, astier13, brout19}). As we have no \RomanST data to validate with, we are left with two choices for test data: inject simulated SNe into images from the \Hubble Wide Field Camera 3 IR channel (\HST WFC3 IR, which has similar filters with a similar pixel scale of $0\farcs128$), or use fully simulated data. We chose to test entirely with simulated data for two reasons. 1) There are a limited number of WFC3 IR visits that have many rotations in several filters (and many visits do not contain a fair sampling of the universe, e.g., they contain a galaxy cluster or a globular cluster). 2) Testing with \HST data will confound \HST calibration uncertainties with forward-model problems. Of particular worry are thermal variations due to the \HST's low orbit \citep{bely93}, PSF spatial variations \citep{andersonwfc3psf}, and detector effects \citep[e.g., ][]{wfc3trap}.

In Section~\ref{sec:intro-pipeline} we very briefly outline the assumptions and requirements for our forward modeling algorithm and how it may interface with the \RomanST pipeline and cosmology analysis. In Section~\ref{sec:mocksimulations} we describe the WFI mock observations which we generate and use to test the forward modeling algorithm. In Section~\ref{sec:forwardmodeling} we explain the forward-modeling assumptions in detail, and in Section~\ref{sec:results} we present and discuss the results. Section~\ref{sec:Summary} summarizes our conclusions.

\section{Forward-Model Inputs and Outputs} \label{sec:intro-pipeline}

\begin{figure}[ht]
    \centering
    \includegraphics[width=\columnwidth]{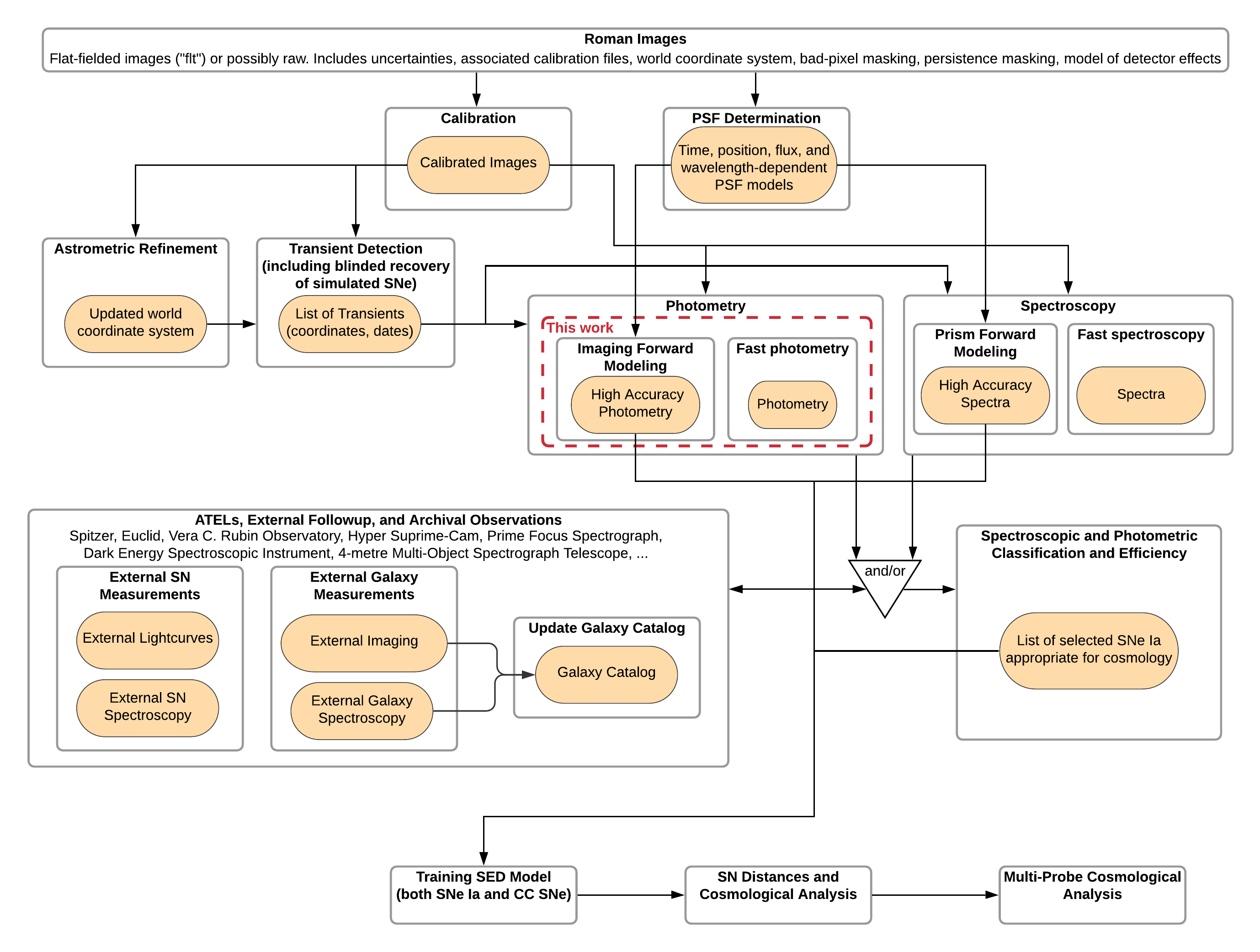}
    \caption{Conceptual \RomanST SN cosmological data-processing flowchart. The red dashed square denotes the photometry component, which is described and validated in this work. We show each step as a single box, but many surveys use more than one semi-independent analyses of the same data as a cross-check \citep[e.g., ][]{guy10}.}
    \label{fig:flowchart}
\end{figure}

Figure~\ref{fig:flowchart} shows a conceptual overview of the \RomanST SN cosmological data-processing flowchart and how this work fits in. The downlinked data will be processed into calibrated images (top of Figure~\ref{fig:flowchart}). \citet{mosby20} discusses the performance of the IR detectors in detail. In short, there are eighteen Teledyne HAWAII 4RG detectors (with 10 micron 4k by 4k pixels) that non-destructively read out every 2.825 seconds. These multiple readouts enable lower read noise than is possible with one read, enable rejecting cosmic-ray hits during a single exposure (visible as jumps in charge vs. readouts), and possibly provide better control of pointing drifts and detector effects \citep{rauscher19}. (To lower the required bandwidth, averaged groups or other linear combinations of readouts will be downlinked.) We assume for this work that the calibration process produces 2D images with known astrometric and photometric calibration. We acknowledge the possibility that we may need to go back earlier in the process (for example by fitting the readouts directly); this will have to be explored with better simulations of the detectors (and ultimately explored with real data).

After the images are calibrated, we assume further processing generates a PSF model. This PSF model will have to take focal-plane position and effective wavelength into account (and possibly temporal or thermal variations as discussed above for \HST). Depending on the linearity calibration \citep{choi20}, it may also have to take flux into account.

The transients in each image will have to be found and assessed. This will require a highly automated process; a 20 deg$^2$ survey with a five-day cadence is equivalent to searching more than 3,000 WFC3 IR pointings per day. \citet{hayden21} demonstrated an automated transient classifier for WFC3 IR data with near-human levels of performance (as noted above, WFC3 IR has similar pixel scale and wavelength coverage), so the search process is feasible, even for these large data sets.

With a series of calibrated images, a PSF model, and transient detections, the forward-modeling code can run. We describe this in more detail in Section~\ref{sec:forwardmodeling}. This step produces calibrated fluxes. The slitless spectroscopy will also need a separate forward-model code appropriate for 3D reconstruction (two dimensions on the sky plus wavelength). This is a fundamentally harder problem (because of the increase in dimensionality, the increase in data volume, and the spatial/spectral degeneracy of a slitless spectrograph observing a complex scene). \citet{ryan18} demonstrates this concept on WFC3 IR data.

The lower half of Figure~\ref{fig:flowchart} outlines the steps involved in going from these calibrated imaging and spectroscopic fluxes to SN distances and cosmological results. We do not elaborate here, as many of the steps will be similar to other surveys \citep[e.g.,][]{scolnic18}.

\section{Simulated Data Generation} \label{sec:mocksimulations}

To create mock observations, we use galaxies from the VELA Cosmological Simulation (\citealt{veladata, 2019ApJ...874...59S}). The dataset spans the cosmic time evolution of 35 galaxies over 10--50 timesteps with cosmological scale factors between 0.05 and 0.5 (redshift 1 to 19), each with approximately 20 viewing angles. The simulated spatially dependent galaxy SEDs are integrated over \RomanST filters (and other existing and proposed observatories), making simulated images. These simulated images are high resolution (oversampled by a factor $\sim$ 15 compared to \RomanST pixels), and are much larger than a PSF (800 by 800 oversampled pixels or $\sim 50$ by 50 native pixels), making them perfect for precision tests of galaxy subtraction. The stellar mass distribution is also similar to SN Ia hosts. We add the time-dependent Type Ia SN fluxes and reproject the high-resolution images onto the WFI detector pixels at different telescope orientations (depending on the position relative to the Sun) for 74 time epochs in a range of 1 year (which is designed to fit well within the two-year planned \RomanST survey).

\subsection{Supernova light curves}

So that our light curves would be reasonably realistic, we generated a sample of SN Ia fluxes using the SALT2-Extended model in the SNCosmo Python package \citep{sncosmo}. SALT2 is a two-parameter (light-curve shape $x_1$ and light-curve color $c$) spectro-temporal model \citep{guy07} that was extended into the UV and NIR with the \citet{hsiao07} template.\footnote{More than one version of SALT2-Extended has been trained. We use the SNCosmo version, not the published one \citep{pierel18}; the SNCosmo version seems to have more accurate rest-frame UV fluxes.} To ensure good sampling of redshift, we assumed a random uniform redshift distribution of the SN sample between $z=0.7$ and 2, instead of following SN rates and cosmological volume. Our assumed cadence is five days \citep{spergel15, hounsell18}. We generated a random time of peak brightness compared to the cadence (so there is not always an epoch right at maximum, nor is maximum always between two epochs). We used a random normal absolute magnitude distribution of $-19.1 - 0.14 x_1 + 3.1 c + \Delta m$ mag, where $x_1$ and $c$ are drawn from random normal distributions centered around 0 with standard deviations of 1 and 0.1, respectively. These are similar to the distributions in, e.g., \citet{scolnic16}, and are intended to span a representative range of signal-to-noise in the simulated imaging. $\Delta m$ is also assumed to be Gaussian, with a standard deviation of $\sqrt{0.1^2 + (0.055 z)^2}$. The $0.055 z$ is the magnitude dispersion due to the weak gravitational lensing of galaxy halos along the line of sight \citep{jonsson10}. Finally, we calculated the integrated flux of the SNe in the five \RomanST bands ($Z087$, $Y106$, $J129$, $H158$, $F184$) over the SALT2 phase range of (-15 to +45 rest-frame days) in 5 observer-frame day steps (an average of 31 epochs).

\subsection{Supernova positions}

We assume that SNe~Ia are distributed following the optical light of the galaxy \citep{anderson15}, and convolve the high resolution $Y106$ VELA image by a Gaussian with a 1 kpc radius before choosing locations to plant SNe.

\subsection{Point spread functions}

We use point-spread functions (PSFs) generated by WebbPSF \citep{webbpsf}. We convolve the PSFs with square $0\farcs11$ pixels with uniform sensitivity. As described in Section~\ref{sec:intro-pipeline}, we assume that detector effects such as count nonlinearity, count-rate nonlinearity, and inter-pixel capacitance have been perfectly calibrated and can be neglected. We also assume that the dependence of the PSFs on the spectral energy distribution (SED) of the source is negligible.\footnote{In practice, one can use an iterative process of generating a PSF, measuring photometry, estimating the SED from a light-curve fit, and re-estimating the PSF \citep{suzuki12}. We neglect this iteration for simplicity. For sufficiently well sampled images, an even simpler approximation suffices: using a single PSF for all SEDs and modifying the filter bandpass instead of the PSF \citep[e.g., ][]{guy10, suzuki12}.} All of these simplifying assumptions are in keeping with our philosophy for this work of focusing on the ``Photometry'' box in Figure~\ref{fig:flowchart}.

\subsection{Reprojection}

\begin{figure}[h!]
    \centering
    \includegraphics[width=\columnwidth]{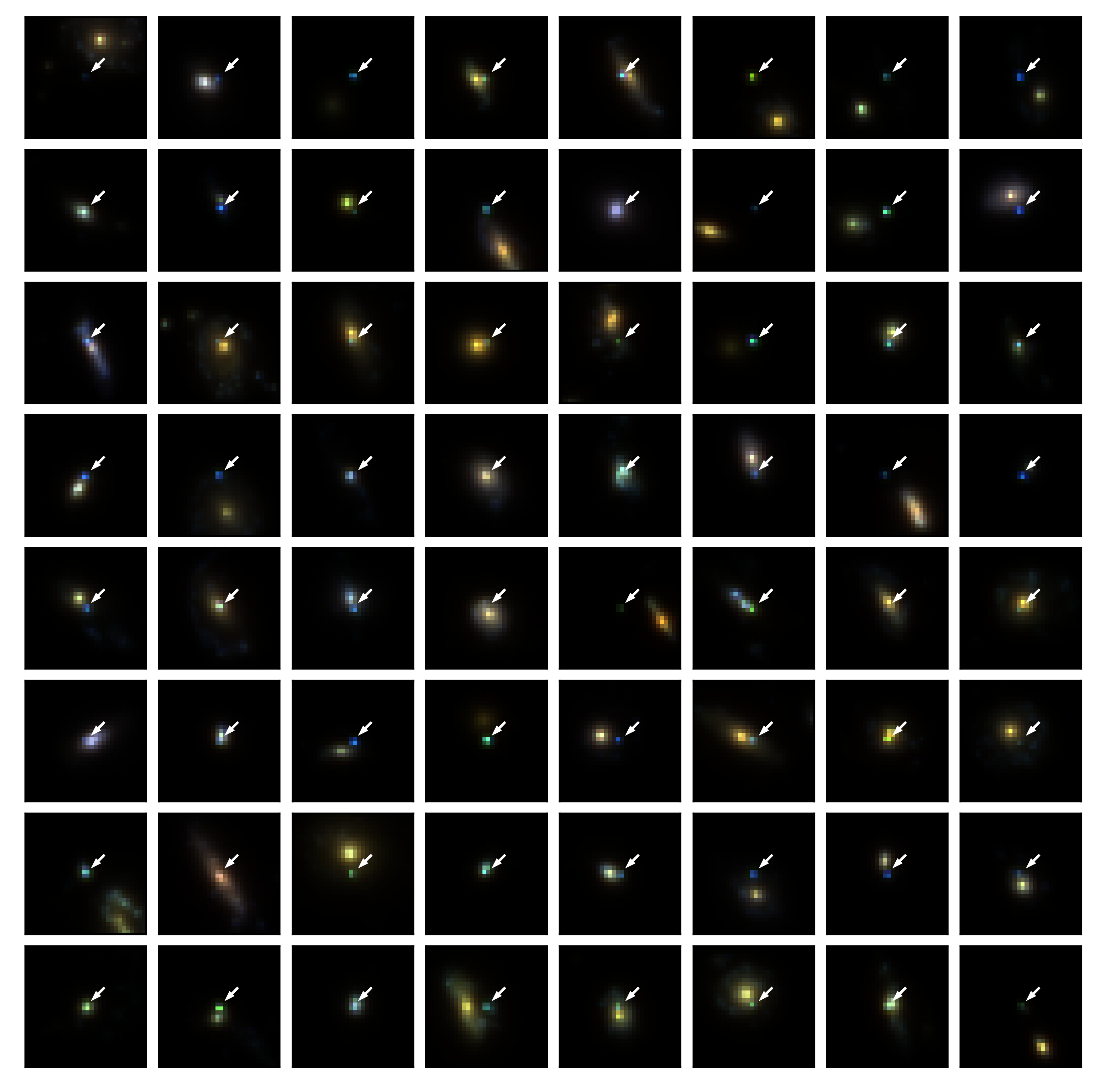}
    \caption{A randomly selected set of postage stamps from our simulations near maximum light for each supernova before the addition of noise. The arrows point out the SN locations. The color channels are $Z087$ (blue), $J129$ (green), and $F184$ (red). The subpixel positions are different for each filter, so we resample the $Z087$ and $F184$ images to match the $J129$ so that the colors align. The VELA galaxies are qualitatively similar to real SN host galaxies (e.g., Figure~2 of \citealt{riess07}). Some of the VELA viewing angles are aligned by angular momentum axis, giving a similar orientation for some galaxies in this figure.}
    \label{fig:postage}
\end{figure}

With a full set of oversampled live-SN and reference images in hand, we use \texttt{Astropy reproject\_exact} to rotate the oversampled data to match the rotation angle for each epoch and give each epoch and filter a random sub-pixel dither offset. This procedure technically convolves the images by each oversampled pixel twice: once when generating the data, and once when rotating the data. We use 30$\xtimes$ oversampled images (pixel doubling the 15$\xtimes$ oversampled VELA images), so this limits the accuracy of our simulations to $\mathcal{O}(30^{-2}) \approx 0.1\%$. We then convolve the rotated image by the PSF and the pixel, and sample it at the native $0\farcs11$ resolution. Figure~\ref{fig:postage} shows a randomly selected set of simulated images (near maximum light for the SNe) before the addition of noise.

WebbPSF defines the PSF as the PSF at exactly the sampled locations. The image-reprojecting code effectively defines the PSF as convolved with the subpixel (in the sense that adding together all the subpixels of a pixel of the PSF should exactly equal the PSF convolved with the pixel). We sample the PSFs at very high resolution ($\sim 100\xtimes$) and integrate over each subpixel to ensure that the pixel convolution is done with high accuracy. Such definition considerations will need to be kept in mind as the \RomanST software stack is built.

\subsection{Sources of Noise}

It is likely that the \RomanST SN survey strategy will consist of tiers \citep{spergel15, hounsell18, rubin21}, with each ``wedding cake'' layer of the survey trading area on the sky against depth. The deepest tier will likely have $\sim 300$ second exposure times and reach redshift~$\sim 2$ with reasonable S/N; the wider tier(s) will have $\lesssim 100$ second exposure times and reach redshift~$\sim 1$ or less. The VELA data are generated only as low as redshift 1, so we will have the most fidelity simulating the deepest tier. To extend the redshift span, we plant simulated SNe as low as redshift 0.7 on the $z=1$ VELA images (the angular scale is only about 12\% different between $z=0.7$ and 1). As higher-redshift SNe have lower contrast against their host galaxy (due to the loss of physical resolution with increasing distance for $z \lesssim 1.6$), the deepest tier presents the most difficult host-galaxy-subtraction problem. Validating the forward-model code in the deep tier will thus validate it in any other tiers as well.

 We assume that this deep tier consists of 300 s exposures in $Z087$, $Y106$, $J129$, and $H158$, and a 600 s exposure in $F184$. It is also possible that the $R062$ filter or the proposed $K213$~band could be used, but these were too recent to be in the Vela simulations, so they are not included here. We briefly note that the $R062$ is comparably undersampled to the $Z087$ after convolution with the pixel, but will have more high-frequency power as the pixel is assumed to be a perfectly sharp square. Thus if we could test $R062$, we may have found worse results than for the $Z087$. Fortunately, $R062$ is currently only planned to be used in the shallower and wider $z \lesssim 1$ tier with $\sim$ 100 second exposure times. This yields an AB magnitude depth almost exactly 1 magnitude shallower than our 300 s $Z087$ exposure. Thus $R062$ can tolerate a factor of a few worse host-galaxy residuals than can the $Z087$. In the other extreme, $K213$ is even better sampled than the data simulated in this work, so it should be straightforward to model and remove the $K213$ galaxy light.

We generate a copy of the images with noise included, but also retain the images without noise for testing. In addition to Poisson noise from the scene (SN + host galaxy), we include zodiacal (based on the model of \citealt{aldering02}) and thermal background  \citep{rubin21} given in Table~\ref{tab:filter}. For read noise, we assume 106 reads for the 300~s exposures and 212 reads for 600~s, using 20 electrons per read with a 5 electron floor, giving values also displayed in Table~\ref{tab:filter} \citep{1993SPIE.1946..395G, 2004PASP..116..352V, 2007PASP..119..768R}.\footnote{As most of the noise in the simulated observations is Poisson noise, we take a simple quadrature sum of the read noise and Poisson noise, without the $\sqrt{6/5}$ scaling on the Poisson noise appropriate for the read-noise-dominated regime. See \cite{fadeyev06} or \citet{kubik16} for a more detailed discussion of possible up-the-ramp estimators.}

Some forward-modeling codes for ground-based data \citep[e.g., ][]{astier13} use only the sky noise (not the source Poisson noise or detector noise) to eliminate biases that would otherwise by caused by by using noisy observations to estimate the Poisson noise. Our photometry is in an even more complex regime: SN Poisson noise, galaxy+sky Poisson noise, and detector noise all matter. We assume that the up-the-ramp readouts have been accurately fit, yielding count-rates with known, Gaussian-distributed uncertainties. As discussed in Section~\ref{sec:intro-pipeline}, we may have to forward model starting with the original detector readouts (which will be read-noise-limited for these faint sources) for satisfactory performance.

If the SN survey takes place over two years, with visits to the SN field(s) every five days, then there will be 146 visits. If a SN goes off at a random epoch, and every image taken after explosion at that location is considered contaminated with SN light, then the number of references will vary up to about 140 (assuming an absolute minimum of $\sim$ 6 epochs required for a good light curve), but be $\sim$ 70 on average. (Including lost epochs due to chip gaps, these numbers are $\sim 124$ and $\sim 62$.) We simulate only a year of data, with an average of 43 reference images per filter, representing a below-average reference set compared to the planned survey, but still relatively representative. As we only simulate one year out of the planned two years of the survey, we do not incorporate missing data, e.g., due to bad pixels or detector gaps, into the simulated images.

\section{Photometry} \label{sec:forwardmodeling}

Following \citet{suzuki12}, we  performed  photometry of  the SNe Ia in the WFI mock observations using a forward-model code (the same one used by \citealt{hayden21} and \citealt{fox21}). This code fits analytic 2D-spline galaxy models (one independent model for each filter) which are convolved with PSFs (including convolution with the pixel) and resampled to match the images. As in \citet{suzuki12}, the modeling uses $0\farcs01$ subpixels (11$\xtimes$ oversampling). Our minimizer of choice is Levenberg-Marquardt \citep{levenberg44, marquardt63}.

The flux of image $i$ on a pixel $x, y$ near the SN location is modeled as the sum of galaxy light $G(\alpha, \delta)$, background light $s_i$, and SN light ($F_i$, which is zero for epochs before explosion or $\sim 1$ rest-frame year after peak). The temporally unchanging galaxy model is evaluated in sky coordinates (right ascension $\alpha$ and declination $\delta$). These must be mapped to the 11$\xtimes$ oversampled pixel coordinates with WCS functions $\mathcal{A}_i: x^{11\xtimes}, y^{11\xtimes} \rightarrow \alpha$ and $\mathcal{D}_i: x^{11\xtimes}, y^{11\xtimes} \rightarrow \delta$. In our code, this is done using \texttt{Astropy all\_pix2world}. These $\alpha$ and $\delta$ values are slightly adjusted with $\Delta \alpha_i$ and $\Delta \delta_i$ values for each image (it remains to be seen if the \RomanST WCS solution will be good enough to avoid these adjustments). The 11$\xtimes$ oversampled galaxy model is convolved with the PSF and the pixel (also 11$\xtimes$ oversampled), then this high-resolution model is sampled every 11 subpixels (including at $x,y$). Thus, 
\begin{equation} \label{eq:galaxymodel}
        G_i(x,y) = G(\mathcal{A}_i (x^{11\xtimes}, y^{11\xtimes}) + \Delta \alpha_i,\ 
    \mathcal{D}_i (x^{11\xtimes}, y^{11\xtimes}) + \Delta \delta_i)
    \otimes \mathrm{PSF}_i(x^{11\xtimes}, y^{11\xtimes})|_{x, y} \;.
\end{equation}
The SN coordinates are also stored in sky coordinates ($\alpha^{\mathrm{SN}}$, $\delta^{\mathrm{SN}}$), but the PSF is in pixels. To convert, we need the inverse of the above WCS transform (\texttt{Astropy all\_world2pix}): $\mathcal{X}_i: \alpha^{\mathrm{SN}},\delta^{\mathrm{SN}} \rightarrow x^{\mathrm{SN}}_i$ and $\mathcal{Y}_i: \alpha^{\mathrm{SN}},\delta^{\mathrm{SN}} \rightarrow y^{\mathrm{SN}}_i$. These are also adjusted with $\Delta \alpha_i$ and $\Delta \delta_i$, giving
\begin{equation}
        x^{\mathrm{SN}}_i,\ y^{\mathrm{SN}}_i = \mathcal{X}_i(\alpha^{\mathrm{SN}} + \Delta \alpha_i, \delta^{\mathrm{SN}} + \Delta \delta_i),\  \mathcal{Y}_i(\alpha^{\mathrm{SN}} + \Delta \alpha_i, \delta^{\mathrm{SN}} + \Delta \delta_i) \;.
\end{equation}
Finally, we can combine both models with the image-dependent, spatially flat sky $s_i$, giving the model $M_i(x,y)$:
\begin{equation}
    M_i(x,y) = G_i(x, y) + \frac{F_i}{A(x, y)} \mathrm{PSF}_i(x - x^{\mathrm{SN}_i},\ y - y^{\mathrm{SN}_i}) + s_i \;
\end{equation}
We assume (as is true for WFC3 IR) that the flat fielding preserves surface brightness, but point-source fluxes are scaled down proportional to the pixel area on the sky $A(x, y)$. Thus the SN model fluxes must also be scaled down to match the data. The galaxy model goes to zero at the edge of the circular fit patch (thus breaking the degeneracy between sky and galaxy light). Note that both the galaxy model and the SN require convolution with the PSF, but we do not insert the SN as a Dirac delta function into the 11$\xtimes$ oversampled galaxy model, as the SN may land with light split between subpixels, broadening the PSF. Finally, we note that other extensions of the formalism (e.g., modifying the PSF shape as a function of image counts, \citealt{choi20}) are straightforward, but we do not consider these here.

There are many possible choices for the galaxy basis functions \citep{thevenaz00}. The general considerations for the galaxy basis functions are that they should be flexible enough to model real galaxies accurately, without being so flexible that the model is poorly constrained, amplifying noise. \citet{holtzman08}, \citet{astier13}, and \cite{brout19} used a grid of squares of constant surface brightness. Smooth parameterizations are also used; \citet{rodet08} used Gaussians and \citet{bongard11} used sinc interpolation of a uniform grid. We want to maintain a smoother model for the undersampled data than a pixelized model or even Gaussians, and want a faster falloff than sinc interpolation (for a smaller modeled patch), so we use 2D splines.  \citet{suzuki12} and \citet{rubin13} also used 2D splines, but added greater flexibility in the galaxy radial direction, giving an overall smoother model to reduce noise for SNe with limited numbers of references. Here, we have many reference epochs, so we simply use 2D splines with a uniform grid. Initially, we experimented with the spline node spacing. In the end, we settled on 2 nodes per PSF FWHM, effectively building an approximately Nyquist-sampled model. In other words, the spline-node spacing for $J129$ was 0.62 pixels or $0\farcs068$ (Table~\ref{tab:filter}).

\section{Results and discussion} \label{sec:results}

\begin{figure}[h]
    \centering
    \includegraphics[width=0.6\textwidth]{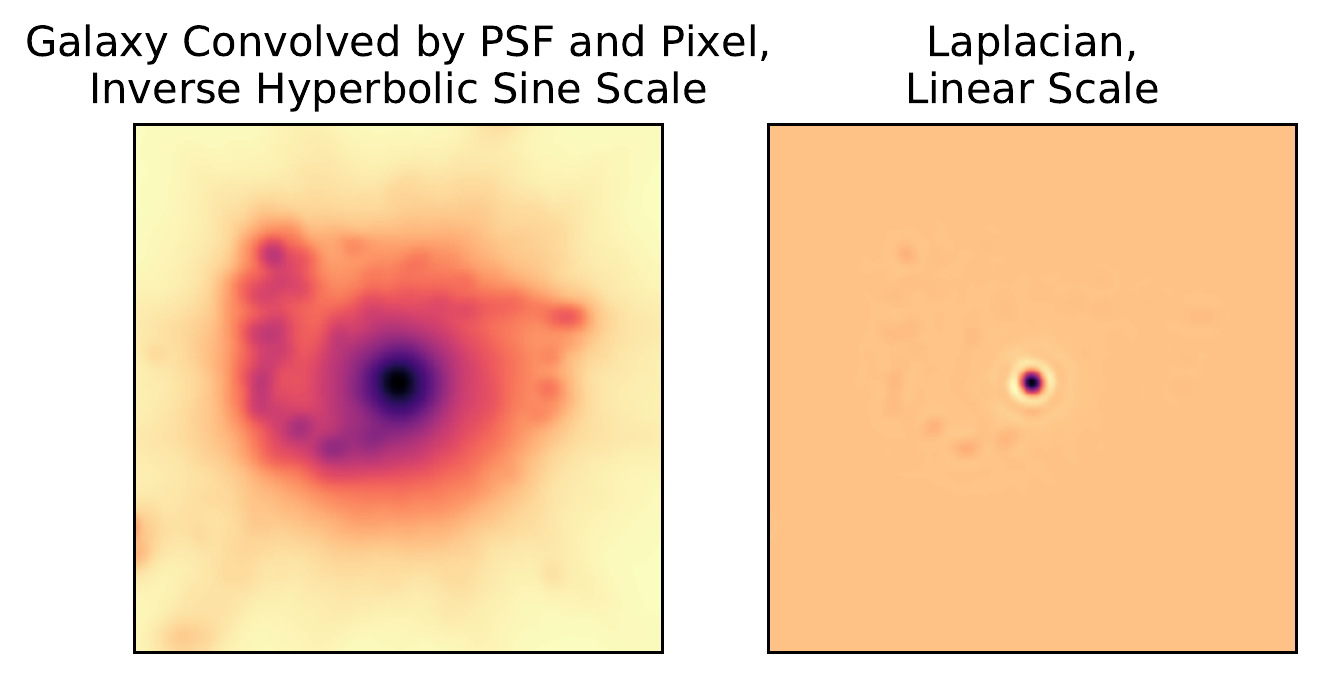}
    \caption{Visualization of the second derivative (Laplacian) for a galaxy. The {\bf left panel} shows the central area around a VELA simulated galaxy, convolved with the PSF and the pixel. To better show faint features, this panel uses inverse hyperbolic sine scaling. The {\bf right panel} shows the second derivative of the left panel. The largest deviations from zero are in a small region of the galaxy around the core. It is these regions that most require spatial flexibility in a galaxy model.}
    \label{fig:galaxylaplacian}
\end{figure}

\begin{figure}[h]
    \centering
    \includegraphics[width =\textwidth]{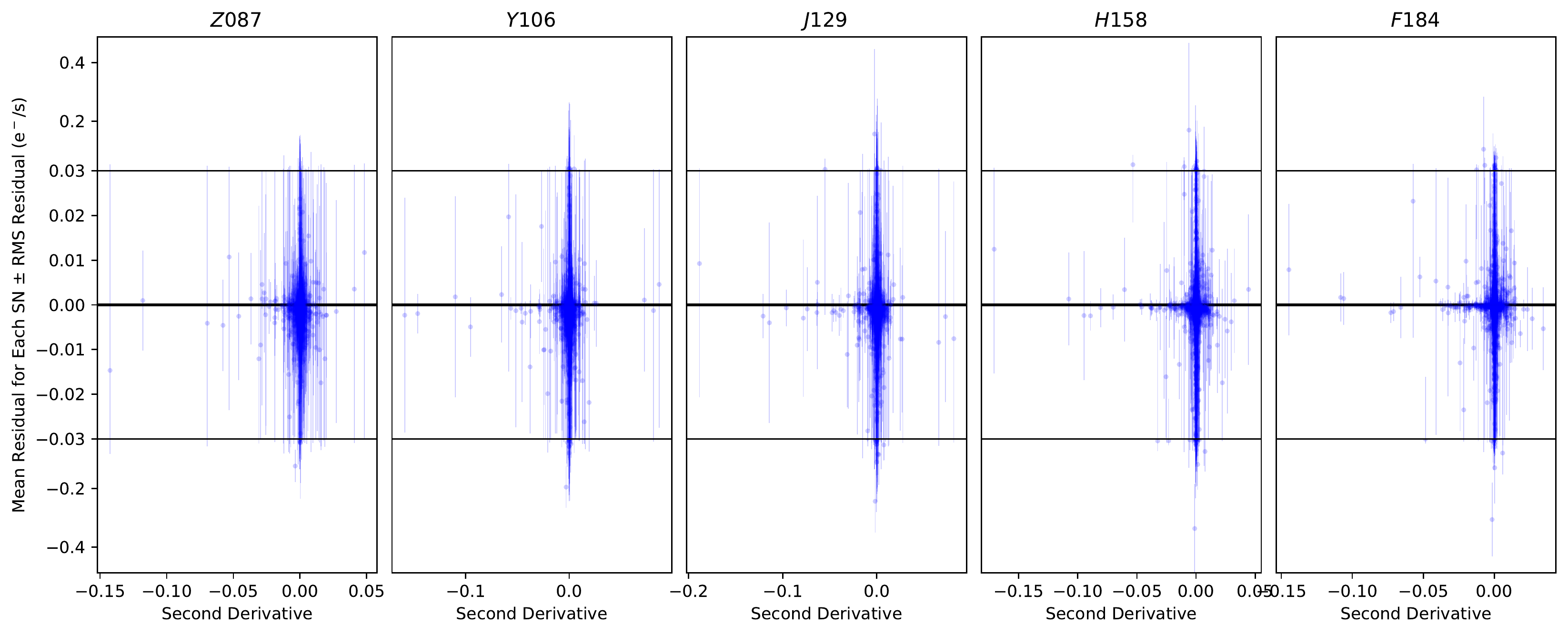}
    \caption{For each band, for each simulated SN, we show the mean flux residual (recovered flux $-$ true flux) over the whole light curve (blue points) plotted against the second derivative of the galaxy flux at the SN location. As illustrated in Figure~\ref{fig:galaxylaplacian}, the second derivative is close to zero over much of the galaxy after convolution with the PSF and the pixel. The error bar on each point shows the RMS across epochs for that SN. To search for any possible trends with the highest sensitivity, we use the forward-model runs on the images that have no noise added. The typical flux at maximum light in this redshift range is 3 e$^-$/s for the four bluest bands and 1.5 e$^-$/s for $F184$, thus these results show very small residuals for almost all the simulated SNe. Note the piecewise-linear scale with an expanded view of $-0.03$ to 0.03 e$^-$/s (roughly $\pm 1\%$ of peak flux). Any trends with galaxy second derivative would indicate that the photometry may need a more flexible galaxy model (e.g., spacing the spline nodes closer together).}
    \label{fig:secderiv}
\end{figure}

We run several sets of analyses to investigate the results of the photometry. Our first result is that our spline-node spacing is sufficient. It is common in ground-based forward modeling to plot results against local galaxy surface brightness (e.g., \citealt{brout19}). However, the galaxy-light Laplacian (second derivative) is the better quantity for most types of modeling errors, as smooth galaxy gradients generally do not cause problems. For ground-based work, the host galaxy is frequently poorly resolved, and the local surface brightness correlates with the Laplacian. Figure~\ref{fig:galaxylaplacian} shows the second derivative for a typical galaxy. For each band, Figure~\ref{fig:secderiv} shows the noise-free residuals plotted against the local second derivative of the host-galaxy light; no trends are seen.\footnote{For a simple comparison, we also perform photometry using a simple image-resampling code for the host-galaxy subtraction. For each image with SN light in it, we take each reference image and resample it to the pixels of that live-SN image. We use a square kernel with a size of 0.3 pixels (frequently called the pixfrac). After subtracting the references, we perform PSF photometry on the subtracted images. We only perform this test with the noise-free images to better examine the differences with forward modeling. This is an extremely simplistic resampling compared to more accurate procedures, e.g., \citet{rowe11} or \citet{fruchter11}. Unsurprisingly, it gives results that are much poorer than the forward model, with large negative slopes visible in all panels indicating over-smoothed galaxy models.}

\begin{figure}[h]
    \centering
    \includegraphics[width =\textwidth]{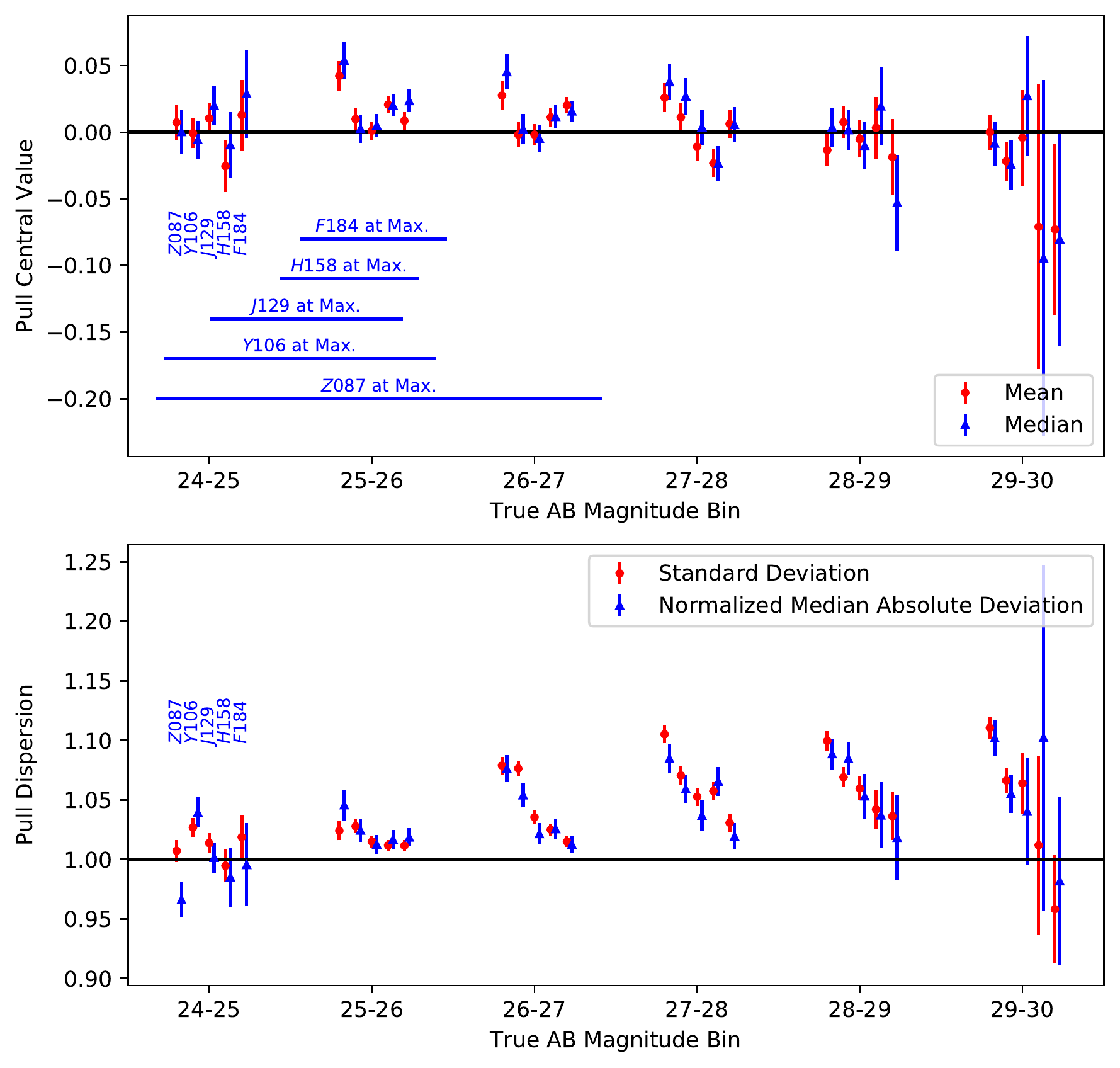}
    \caption{Summary statistics for pulls: (recovered~flux $-$ true~flux)/recovered~flux~uncertainty. The results are binned in AB magnitude and separate results are shown for each filter (left to right is $Z087$ to $F184$). The {\bf top panel} shows central values (mean with red dots, median with blue triangles). In general, there is no evidence for biases (offsets from zero) except in the $Z087$ filter in the middle of the magnitude range ($\sim$ 25--28). We also show the 16th to 84th percentile of SN fluxes at maximum for our simulated sample. The {\bf bottom panel} shows dispersions (standard deviation with red dots, the normalized median absolute deviation with blue triangles). The uncertainties on the NMAD are computed with bootstrap resampling. If all uncertainties are correct and Gaussian, the dispersion values should be unity. There is evidence of mildly underestimated (by $\lesssim$ 10\%) flux uncertainties in bluer filters for magnitudes fainter than $\sim 26$.
    \label{fig:pullvsmag}}
\end{figure}

Next, we search for biases and check the uncertainties by plotting distributions of pulls: (recovered~flux $-$ true~flux)/(recovered~flux~uncertainty). Figure~\ref{fig:pullvsmag} shows summary statistics, binned in true AB magnitude:  zeropoint~$-$~$2.5 \log_{10}$(true~flux). In general, the forward-model code has better performance in the redder filters with better sampling. The mildly underestimated ($\lesssim$ 10\%) flux uncertainties in bluer filters at faint magnitudes might plausibly be due to the small galaxy-subtraction residuals shown in Figure~\ref{fig:secderiv}.

\begin{figure}[h]
    \centering
    \includegraphics{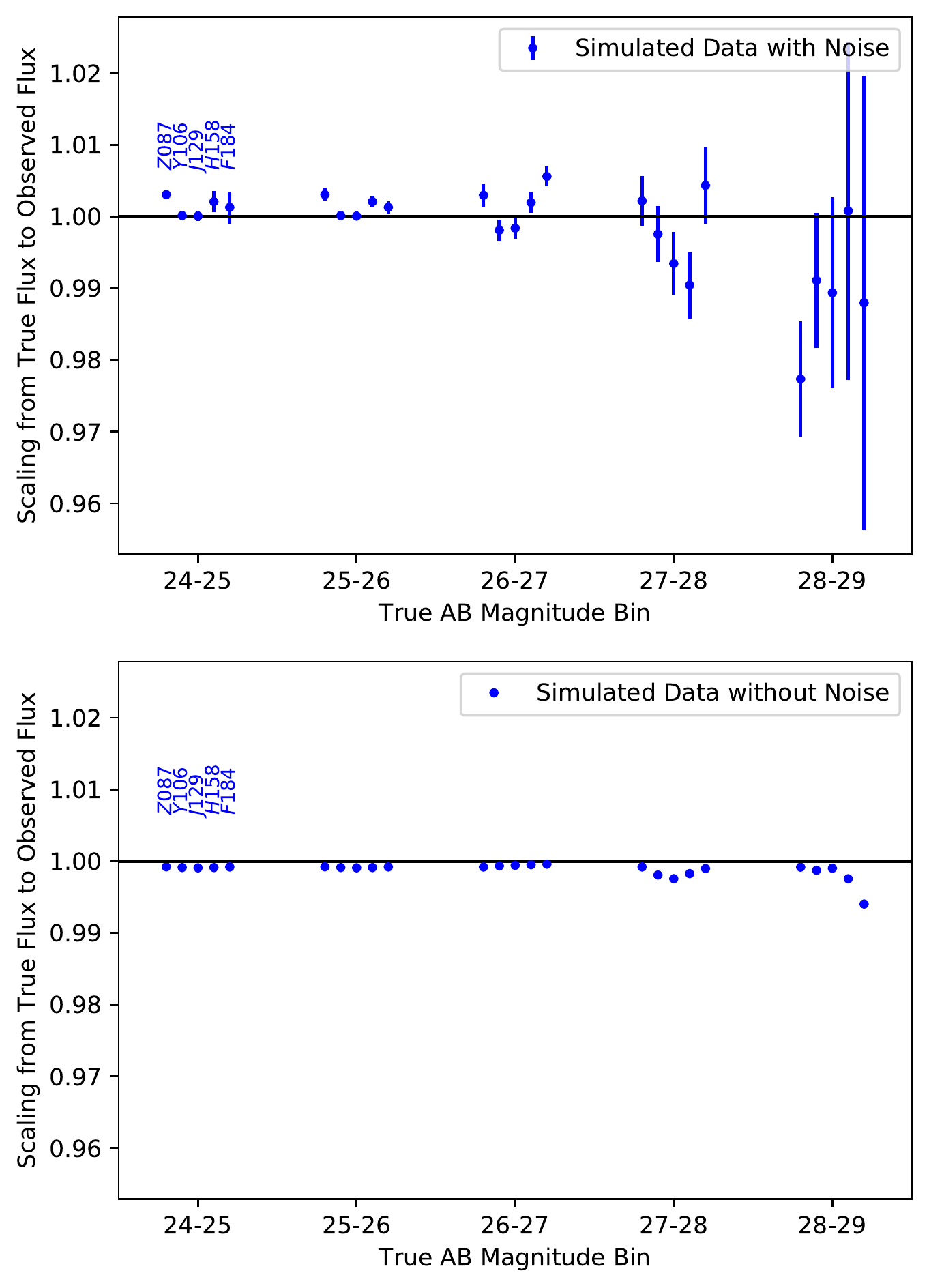}
    \caption{Average scaling on true flux to match observed flux, binned by AB magnitude and separated by filter. The {\bf top panel} shows the results from the images with noise added. A small consistent bias is seen in the $Z087$ filter (leftmost point in each bin). The {\bf bottom panel} shows the results from images without noise added. The brightest three magnitude bins show a slight bias (probably due to the way the data were generated as discussed in Section~\ref{sec:mocksimulations}). The faintest magnitude bins show a larger (but still small) bias.
    \label{fig:linearvsmag}}
\end{figure}

Figure~\ref{fig:linearvsmag} shows observed flux regressed on true flux. We use both images with noise (top panel) and images without noise (bottom panel). For the images without noise only small biases ($\sim 1$~mmag or 0.1\%) are seen until faint magnitudes. These are likely caused by the accuracy of the image reprojection (discussed in Section~\ref{sec:mocksimulations}). At fainter magnitudes, the accuracy degrades, possibly due to the slight galaxy-subtraction residuals seen in Figure~\ref{fig:secderiv}). For the results including noise, $\sim 2$~mmag biases are seen in $Z087$ but the other filters are generally consistent with unity mean scaling between true and observed fluxes.

\begin{figure}[h]
    \centering
    \includegraphics{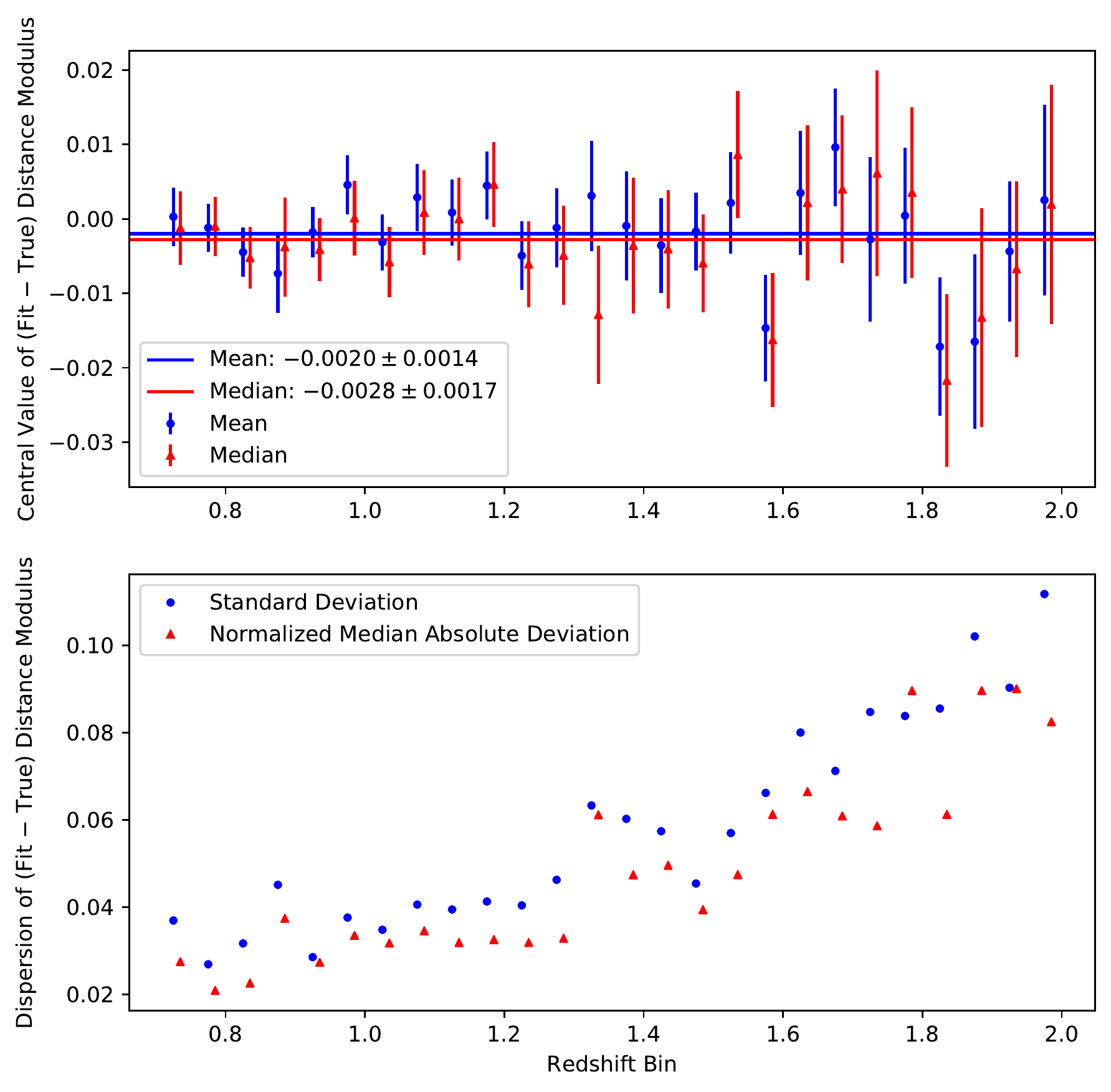}
    \caption{Summary statistics on recovered distance modulus compared to true distance modulus for each SN. The {\bf top panel} shows the mean (blue circles) and median (red triangles) in bins of redshift. The blue and red lines show the result over all redshifts. No strong evidence of bias is seen. The {\bf bottom panel} shows the dispersion in each bin (blue dots for the mean, and red triangles for the normalized median absolute deviation). The increase as a function of redshift is due to a combination of the lower signal to noise, and the loss of red rest-frame wavelength coverage.}
    \label{fig:distancemodulus}
\end{figure}

\begin{figure}[h]
    \centering
    \includegraphics[width=0.8 \textwidth]{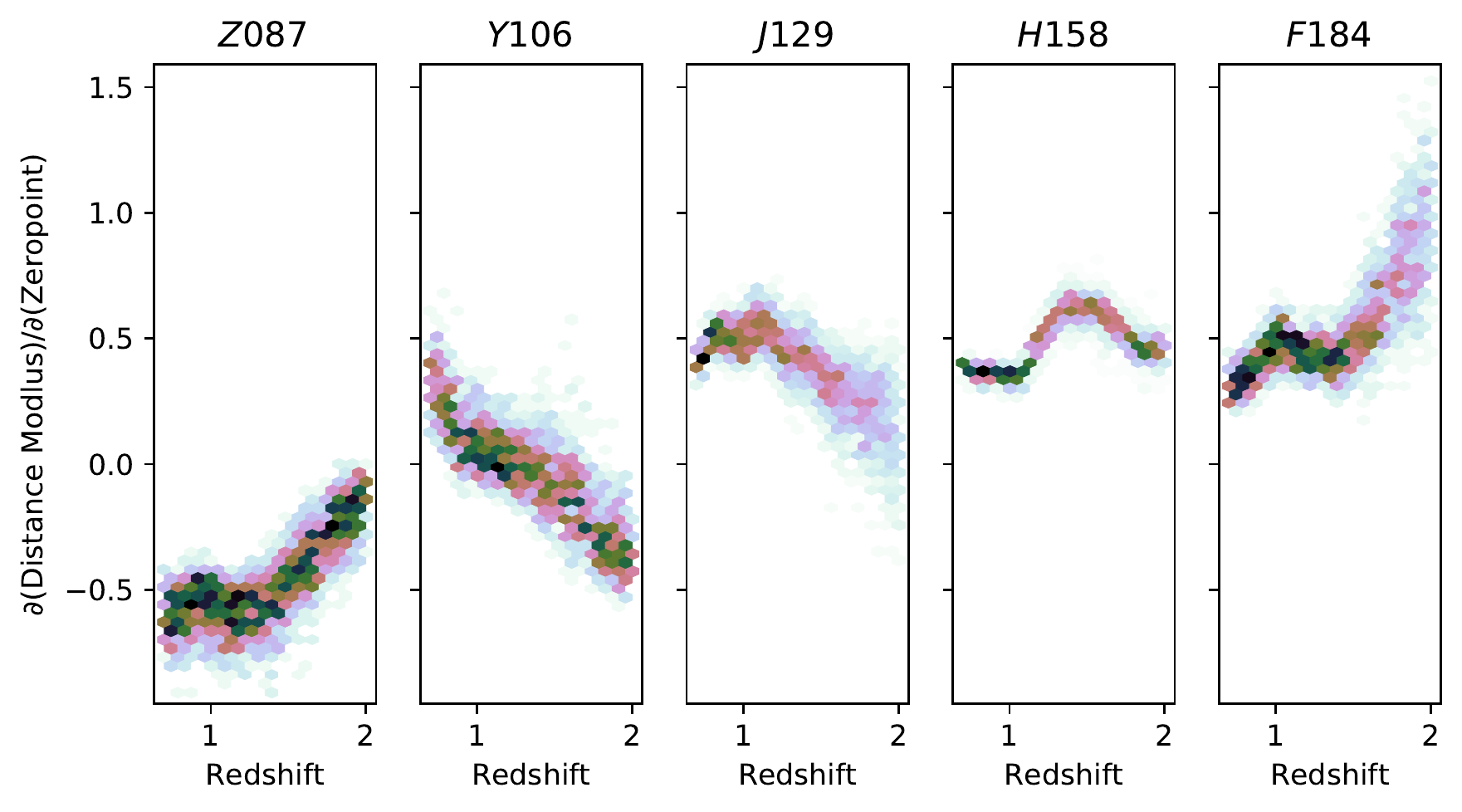}
    \caption{Sensitivity of the distance modulus of each SN to the calibration of each filter, plotted as a function of the redshift of each SN. As in \citet{amanullah10}, this is computed by scaling the calibration for each filter in turn, refitting the SN with SALT2, and computing a new distance modulus. The distance modulus difference divided by the size of the shift in magnitudes gives the derivative. We do not propagate these calibration changes into the training of SALT2, which will change the results in detail \citep{guy10}. The sum of all the derivatives should be 1 (moving the calibration of each filter by 1 magnitude should move the distance modulus by 1 magnitude), and \citet{amanullah10} note that unstable light-curve fits frequently reveal themselves as deviations from 1. We exclude three such SNe from this plot which show up as outliers. For the five-band light-curves considered in this work, the sensitivity of the distance moduli to the calibration of any one filter is $\lesssim 1$. Limiting the wavelength range or number of bands will significantly increase the sensitivity to calibration, resulting in much larger values than those shown here. For example, with just rest-frame $B$ and $V$ data (and using 3.1 for the slope of the color-magnitude relation), the distance moduli scale as $m_B - 3.1 (m_B - m_V) = 3.1 m_V - 2.1 m_B$.
    \label{fig:dmudzp}}
\end{figure}

Finally, we fit light curves using SALT2-Extended; Figure~\ref{fig:distancemodulus} shows these results. Any biases seem to be at the few mmag level or smaller. Figure~\ref{fig:dmudzp} shows the sensitivity of our distance moduli to the calibration of each filter as a function of redshift. Our constraints on the cosmological bias are thus expected from the accuracy with which we recover the light-curve fluxes, but this test still uniquely measures any correlated effects of host-galaxy subtraction on the full light curves.

\section{Summary} \label{sec:Summary}

We validate a forward-model code for performing SN photometry in simulated undersampled images for the \RomanST transient survey. As there are no real images to inject simulated SNe into, we use the VELA simulated galaxy images, which are generated in the \RomanST filters over a similar redshift range as the SN survey. We create 762,570 simulated postage stamps around the locations of 2,061 simulated SNe ($\sim 10\%$ of the anticipated full survey). We describe the assumptions of our forward-model code and validate those assumptions first with noise-free images, and then with images that have noise added. Finally, we fit our simulated light curves and show that we can recover SN distance moduli with biases limited to less than a few mmags. The forward-model code has been released on Zenodo \citep{zenodo21}.

\acknowledgments

We thank Susana Deustua and the anonymous referee for careful feedback. This work was supported by NASA through grant NNG16PJ311I (Perlmutter \RomanST Science Investigation Team). The technical support and advanced computing resources from the University of Hawai`i Information Technology Services Cyberinfrastructure are gratefully acknowledged. This work was also partially supported by the Office of Science, Office of High Energy Physics, of the U.S. Department of Energy, under contract no. DE-AC02-05CH11231. This research used resources of the National Energy Research Scientific Computing Center, a DOE Office of Science User Facility supported by the Office of Science of the U.S. Department of Energy under Contract No. DE-AC02-05CH11231.

\software{
Astropy \citep{astropy},
Mathematica \citep{Mathematica},
Matplotlib \citep{matplotlib}, 
Numpy \citep{numpy}, 
Python \citep{python3}, 
SciPy \citep{scipy},
SNCosmo \citep{sncosmo}
}

\end{document}